\documentclass[journal]{IEEEtran}

\usepackage{cite}
\usepackage{url}

\usepackage{graphicx} 
\usepackage[usenames]{xcolor}
\usepackage[normalem]{ulem}

\newcommand{\yet}[1]{\textcolor{red}{#1}}

\ifCLASSOPTIONcompsoc
\usepackage[caption=false,font=normalsize,labelfon
t=sf,textfont=sf]{subfig}
\else
\usepackage[caption=false,font=footnotesize]{subfi
g}
\fi
\usepackage[cmex10]{amsmath}
\usepackage{amsthm,amssymb}
\usepackage{algorithm,algorithmic}

\usepackage[]{footmisc}
\begin{document}
\title{Attribute-Based Access Control for Smart Cities: A Smart Contract-Driven Framework}
\author{Yuanyu Zhang, \IEEEmembership{Member, IEEE},
		Mirei Yutaka,
		Masahiro Sasabe, \IEEEmembership{Member, IEEE},
		and Shoji Kasahara, \IEEEmembership{Member, IEEE}
\thanks{A preliminary version of this paper appeared in \cite{yutaka19using}, presented at the IEEE Global Communications Conference (GLOBECOM), Hawaii, USA, in 2019.} 
%This paper significantly differs from the conference version from many aspects including the title, Abstract, Introduction and Related Work sections. Also, extensive new experimental results are added in a new section, i.e., Section \ref{sec:evaluation} Cost Evaluation, to justify the novelty of this version.}}
\thanks{Y. Zhang, M. Yutaka, M. Sasabe and S. Kasahara are with the Graduate School of Science and Technology, 
        Nara Institute of Science and Technology, Ikoma, Nara, Japan. E-mail:yutaka.mirei.yj7@is.naist.jp, \{yy90zhang, m-sasabe,kasahara\}@ieee.org.}	
}% <-this % stops a space

\maketitle

\begin{abstract}
Efficient and reliable access control in smart cities is critical for the protection of various resources for decision making and task execution. Existing centralized access control schemes suffer from the limitations of single point of failure, low reliability and poor scalability. This paper therefore proposes a distributed and reliable access control framework for smart cities by combining the blockchain smart contract technology and the Attribute-Based Access Control (ABAC) model. The framework consists of one Policy Management Contract (PMC) for managing the ABAC policies, one Subject Attribute Management Contract (SAMC) for managing the attributes of subjects (i.e., entities accessing resources), one Object Attribute Management Contract (OAMC) for managing the attributes of objects (i.e., resources being accessed), and one Access Control Contract (ACC) for performing the access control. To show the feasibility of the proposed framework, we construct a local private Ethereum blockchain system to implement the four smart contracts and also conduct experiments to evaluate the monetary cost as well as to compare the proposed framework with an existing Access Control List (ACL)-based scheme. The experimental results show that although the proposed scheme consumes more money than the ACL-based scheme at the deployment stage, it introduces less monetary cost during the system running especially for large-scale smart cities. 

\end{abstract} 
\begin{IEEEkeywords}
Smart cities, attribute-based access control, blockchain, smart contract. 
\end{IEEEkeywords}

\IEEEpeerreviewmaketitle

\section {Introduction}\label{sec:intro}
\subsection{\yet{Research Problem}}
\IEEEPARstart{T}{hanks}  to the rapid development of communication and manufacturing technologies, the Internet of Things (IoT) has been expanding at an unprecedented speed  in last decades \cite{gubbi2013iot}. 
Smart city, which brings intelligence to various aspects of our life such as healthcare, logistic and transportation, has been recognized as a representative IoT application. 
To actualize the vision of smart cities, a huge amount of resources (e.g., sensors, actuators and data) need to be deployed in cities to facilitate decision making and task execution. 
These resources help make our life increasingly intelligent and convenient, while their vulnerability to unauthorized access puts our property and safety in danger \cite{alaba2017iot, Mirai, webcam}. 

\yet{Access control has been regarded as an effective solution to prevent unauthorized resource access and has found various applications in smart cities \cite{Sookhak2019IEEECST,Eckhoff2018IEEECST}.
One typical and specific example is the access control in smart healthcare, where access to electronic health records of citizens is strictly controlled to avoid leakage and misuse of private information \cite{Algarni2019IEEEAccess}. 
Another example is the access control in smart building, where electronic locks and keys form a smart access control system \cite{locken}. 
Each smart key stores permissions to open a list of locks.
Apart from smart healthcare and smart building, access control has also found applications in other systems, like smart parking and logistics \cite{Gharaibeh2017IEEECST}.
To summarize, access control plays a significantly important role in securing various aspects of smart cities.}

\yet{Each sub-system in smart cities, like health, education and transportation, has its own access control system.
As a result, it is extremely difficult to implement conventional centralized and unified access control in smart cities, as the centralized access control server can easily become a bottleneck \cite{yavari2017scalable, liu2017anaccess, yuan2005attributed, hwang2005TKDE, hwang2009TKDE, li2011privacy}.
Also, the server itself turns out to be a single point of failure and would sabotage the whole access control system once it is destroyed by man-made or natural disasters.
More importantly, centralization renders the access control system vulnerable to attacks, since attackers may easily compromise the whole system by attacking the server only. 
As per above, access control in smart cities needs to be \emph{decentralized} and \emph{trustworthy} (i.e., robust to attacks) to cope with the large-scale and distributed nature. }

\yet{Apart from being decentralized and trustworthy, access control in smart cities also needs to be \emph{dynamic}. 
This means that access control systems should be able to automatically adapt to the dynamically changing access context information, like time, location and situations.
For example, normally permitted access to a smart building should be prohibited when the building is in a fire accident. 
Access to an electronic health record should be restricted to requests from only a certain IP address scope.
If the IP address is beyond the scope, the access request must be denied.}

\yet{Another critical requirement of access control in smart cities is that it must be sufficiently \emph{fine-grained} to achieve high flexibility and accuracy. 
For example, in a smart healthcare scenario, a general policy may be "Doctors can view the record of Patient A."
However, this policy is too coarse to accurately control the access requests, because there are different types of doctors and maybe only a certain type (e.g., psychiatrists) is allowed to view the record, or even more strictly, only the psychiatrists in charge of Patient A can view the record.
Smart cities usually contain various objects (i.e., resources to be accessed) with different access control requirements and subjects (i.e., entities sending access request) with different attributes. 
Thus, \emph{fine-grained} is one of the must-satisfy requirements for the access control in smart cities.}

\yet{In light of the above, this paper aims to design a \emph{decentralized}, \emph{trustworthy}, \emph{dynamic} and \emph{fine-grained} access control scheme to prevent unauthorized access to IoT resources in smart cities.}

\subsection{\yet{Research Methods}}
\yet{Although traditional distributed access control schemes (e.g.,  \cite{Hernandez-Ramos2013} and \cite{Sciancalepore2018}) are able to meet the distributed requirement, they can hardly guarantee the trustworthiness. 
This motivates researchers to apply the highly promising blockchain technology for decentralized and trustworthy access control \cite{dukkipati2018decentralized, wang2019anattribute, maesa2019ablockchain, yutaka19using, dorri2017blockchain, maesa2017blockchain, zhu2018tbac, ouaddah2017access, zhang2018smart, tanzeela2020data, xu2018blendcac, nakamura19capbac, albreiki2019decentralized, lyu2020sbac, jasonpaul2018rbacsc, hao2019multi, ding2019anovel, yu2020enabling, suciu2019attribute}.
Blockchain, at its core, is a distributed database managed over a Peer-to-Peer (P2P) network \cite{Bitcoin,Ethereum}. 
The integrity and authenticity of data are verified by the majority of the network nodes, thus creating a tamper-resistant database.
Current blockchain realizations like Ethereum \cite{Ethereum} also support distributed and tamper-resistant computing through the introduction of smart contract functionality.
Thus, by storing access control policies and implementing access control logic on the blockchain, we can achieve decentralized and trustworthy access control for smart cities.}

\yet{Classical access control models mainly include Access Control List (ACL), Role-Based Access Control (RBAC), Capability-Based Access Control (CapBAC) and Attribute-Based Access Control (ABAC), among which ABAC is the most promising one to achieve dynamic and fine-grained access control.
This is because ABAC introduces context information and also the attributes of subject and objects into its access control policies.
By adding more subject attributes, object attributes and context information into policies, we can greatly improve the dynamicity and granularity of ABAC.
In addition to dynamicity and fine granularity, ABAC offers several other advantages in smart cities. 
First, ABAC enables more accurate access control in smart cities than other models by including sufficient attributes.
Second, ABAC enables access control from a larger set of subjects to a larger set of objects without specifying individual relationships between each subject and each object.
Third, ABAC reduces the burden of maintenance, as access policies can be changed by simply changing the attribute values without the need to change the underlying subject-object relationships.  
}

\yet{Motivated by the benefits of the blockchain technology and ABAC model, we propose a novel blockchain-based ABAC framework in this paper.
Each ABAC policy is responsible for the access control between a set of objects and a set of subjects, i.e., \emph{many-to-many} access control.
The core idea of the proposed framework is to store the attributes, policies and access control logic in smart contracts. 
More specifically, the proposed framework consists of one Policy Management Contract (PMC), one Subject Attribute Management Contract (SAMC), one Object Attribute Management Contract (OAMC) and one Access Control Contract (ACC). 
The SAMC, OAMC and PMC are used to store and manage (e.g., update, add and delete) the attributes of subjects, the attributes of objects and the ABAC policies, respectively. 
When receiving the access request from a subject, the ACC retrieves the corresponding policy, subject attributes and object attributes from the PMC, SAMC and OAMC, respectively, to perform the access control. 
A prototype system on a local private Ethereum blockchain network was constructed to demonstrate the feasibility of the proposed framework. 
Extensive experiments were conducted to evaluate the monetary cost of the proposed framework. 
Comparisons with the existing ACL-based framework in \cite{zhang2018smart} are also conducted to show the superiority of our framework.}

\subsection{\yet{Contributions and Innovation}}

Some recent work has been done to investigate blockchain-based ABAC. 
The authors in \cite{dukkipati2018decentralized} proposed a smart contract-based ABAC scheme, which stores the URLs of policies on the blockchain while leaving the policies themselves in external databases. 
To access an object, subjects send the URL of the corresponding policy to a smart contract, which then retrieves the policy from the external databases and performs the access control. 
Although the storage overhead can be reduced to some extent by storing only the policy URLs on the blockchain, the policies face \yet{a} high risk of being falsified, which may result in untrustworthy access control. 
Besides, the authors provided no implementations to verify the feasibility of the scheme. 
To address this issue, the authors in \cite{wang2019anattribute} proposed a new ABAC scheme, which stores the policies and attributes in smart contracts to achieve trustworthy access control. 
In \cite{maesa2019ablockchain}, instead of storing policies in smart contracts, the authors transfer each policy into a smart contract. 
Although these two schemes demonstrate the feasibility of blockchain-based ABAC schemes, they follow the same idea of \emph{one-to-many} access control, i.e., each policy is associated with one object and responsible for the access requests of many subjects. 
This may incur a huge burden of policy management, especially in large-scale IoT systems like smart cities. 

\yet{Compared with existing blockchain-based ABAC schemes, the proposed ABAC framework have the following major contributions and innovation. 
\begin{itemize}
\item Compared with \cite{dukkipati2018decentralized}, the proposed ABAC framework can greatly improve the trustworthiness by storing ABAC policies and attributes instead of their URLs on the blockchain.
In addition, the proposed ABAC framework follows the standard XACML architecture \cite{standard2013extensible}, while the framework in \cite{dukkipati2018decentralized} does not.
More importantly, we provide implementations and experiments to illustrate the feasibility and performances of the proposed framework.
\item The idea of the proposed ABAC framework differs greatly from the one in \cite{maesa2019ablockchain}, where each policy has to be hardcoded into a smart contract.
This incurs a huge burden of coding when a large number of policies need to be added.
In our framework, this can be easily realized by simply adding the policies into the blockchain without any coding.
\item The most relevant framework is the one in \cite{wang2019anattribute}, which also follows the idea of storing policies, attributes and logic on the blockchain.
The main difference is that each policy in \cite{wang2019anattribute} enables only one-to-many access control, while our policies enables many-to-many access control.
This can greatly reduce the burden of policy management and also storage cost, especially in smart cities with a huge amount of subjects and objects.
\end{itemize}
}

The remainder of the paper is organized as follows. Section \ref{sec:related_work} introduces the related work and Section \ref{sec:proposed_scheme} presents the proposed framework. We provide the implementation in Section \ref{sec:implementation}, the monetary cost evaluation in Section \ref{sec:evaluation} and finally conclude this paper in Section \ref{sec:conclusion}.
 
%%%%%%%%%%%%%%%%%%%%%%%%%%%%%%%%%%%%%%%%%%
\section{Related Work}\label{sec:related_work}
\subsection{Blockchain Technology}
The blockchain technology, the core of modern cryptocurrency systems like the Bitcoin \cite{Bitcoin}, is essentially a distributed database managed over a Peer-to-Peer (P2P) network. 
All peers in the network maintain the same copy of the blockchain and synchronize to update the blockchain. 
A blockchain consists of a sequence of blocks, each containing a collection of transactions recording the remittance information (e.g., sender, receiver and amount). 
These blocks are chained together by each storing the cryptographic hash of its previous block. 
The hash of a block is generated by a process called mining, which requires a huge amount of calculations. 
Suppose an attacker manages to tamper with the transactions in a certain block, he has to re-calculate the hash values of this block and also its subsequent blocks. 
This is considered impossible in general and thus makes blockchain a tamper-resistant database. 

Although originally developed as a distributed database, blockchain has currently advanced to a distributed computing platform thanks to the emergence of a new functionality called smart contract. 
The most representative realization of such novel blockchain is the Ethereum \cite{Ethereum}.  
An Ethereum smart contract is an executable program, which, like transactions, is also stored (indirectly) on the blockchain. 
A smart contract consists of variables as its states and functions called Application Binary Interfaces (ABIs) to view and change the states \cite{SC}. 
To execute a smart contract, a transaction needs to be fired and broadcast to the P2P network.
 All peers receiving this transaction will execute the smart contract to ensure the validity of execution results. 
 Thus, by implementing the smart contract functionality, the blockchain technology can further achieve distributed and tamper-resistant computing.
 
\subsection{Blockchain-based Access Control}
To deal with the access control issue in a smart home, a Bitcoin-like blockchain was adopted in \cite{dorri2017blockchain} to design an  ACL-based access control scheme. 
A blockchain was implemented locally in each home to maintain an ACL, where each entry corresponds to an internal object and records the allowed access rights of a subject (internal or external). 
The access control inside each home is performed by an internal miner, which acts as the gateway to receive access requests from subjects.
However, the existence of the miner results in centralized access control inside each home. 
Besides, the mining process is eliminated by the miner, which makes tampering with the ACL possible and thus results in untrustworthy access control. 
Based on the Bitcoin blockchain, the authors in \cite{maesa2017blockchain} proposed an ABAC scheme, which stores ABAC policies in Bitcoin transactions and serves as a policy repository for existing ABAC solutions. 
To update the policy inside a certain transaction, administrators can append to the transaction a new transaction with \yet{updated} information.
In addition, administrators can simply spend the coins contained in a transaction to delete the policy inside the transaction. 
The authors  in \cite{zhu2018tbac} proposed a similar Bitcoin-based ABAC scheme, while the policies are encrypted for privacy and security, which is different from \cite{maesa2017blockchain}.
A Bitcoin-based CapBAC scheme was designed in \cite{ouaddah2017access}, which stores capability tokens (i.e., special data structures that record the assigned access rights of a certain subject to one or more objects) in Bitcoin transactions.
A subject can transfer his/her capability token to another subject through transactions, which is similar to the transfer of bitcoins.
When accessing an object, subjects must prove their ownership of the corresponding capability tokens (i.e., the access rights) to the object owner.

The above schemes are all based on the Bitcoin-like blockchains, while smart contract-based access control has attracted more attention recently.
For example, the authors in \cite{zhang2018smart} applied Ethereum smart contracts to store ACLs in order to propose an ACL-based access control framework.
In the framework, each subject-object pair uses one smart contract to implement the related access control. 
When accessing an object, a subject sends a transaction, which contains the required access information, to execute the corresponding smart contract. 
After the execution, the smart contract will automatically return the results (denial or permission) to both the subject and object. 
However, since one contract is responsible for the access control of only one subject-object pair, this scheme suffers from heavy monetary cost of deploying contracts, especially in large-scale IoT systems.
The authors in \cite{tanzeela2020data} extended the above framework with slight modification. 
The authors in \cite{xu2018blendcac} proposed a CapBAC scheme, which applies a smart contract to store the capability tokens and capability delegation tokens that record the delegation information among the subjects. 
These tokens are managed in a tree form and serve as tamper-resistant references to help object owners decide whether a subject has certain access rights.
This scheme was extended in \cite{nakamura19capbac}, where the authors used a delegation graph to replace the delegation tree in \cite{xu2018blendcac}. 
In addition, the authors designed only one type of tokens rather than two types. 
The authors in \cite{albreiki2019decentralized} also proposed a CapBAC-like scheme to manage access control for data sharing in IoT systems, where oracles are used to connect blockchain, data hosts, and users for data accessing. 
Another CapBAC-like scheme was proposed in \cite{lyu2020sbac} for handling the access control in information-centric networks. 
In \cite{jasonpaul2018rbacsc}, an RBAC scheme was designed, where a smart contract was deployed to maintain the roles assigned to each user in an RBAC model, such that any service provider can verify the users' ownership of roles when providing services. 
An ABAC scheme was proposed in \cite{dukkipati2018decentralized}, where the URL links of ABAC policies are stored on the blockchain.
In addition, a smart contract is deployed to receive access requests from the subjects and then perform the access control. 
However, the main limitation of this scheme is storing the policies and attributes in external databases, which makes them vulnerable to tampering attacks.
As a result, the trustworthiness of the policies and the attributes cannot be guaranteed. 
Besides, the authors did not provide implementations, and thus the feasibility of the proposed scheme is not clear.
Another ABAC scheme was also proposed  in \cite{hao2019multi}, while the object attributes were not considered and only one-to-many access control was realized. 
Other ABAC schemes were also proposed in \cite{wang2019anattribute, maesa2019ablockchain, yutaka19using} as introduced in Section \ref{sec:intro}.

Access control schemes based on other blockchain realizations have also been designed. 
For example, an ABAC framework based on the permissioned Hyperledger Fabric blockchain was proposed in \cite{ding2019anovel}, while, different from the schemes in \cite{yutaka19using, wang2019anattribute}, only the attributes are stored on the blockchain and no smart contracts are used for processing access requests. 
The authors in \cite{yu2020enabling} combined  attribute-based encryption and  blockchain to  propose another ABAC-like scheme based on a multi-layer blockchain architecture. 
In \cite{suciu2019attribute}, a conceptual design of blockchain-based ABAC was provided, while the authors provided no implementations.

\section{Proposed ABAC Framework}\label{sec:proposed_scheme}
\subsection{Smart Contract System}\label{sec:sc_system}
\begin{figure}[!t]
 \centering
 \includegraphics[width=\columnwidth]{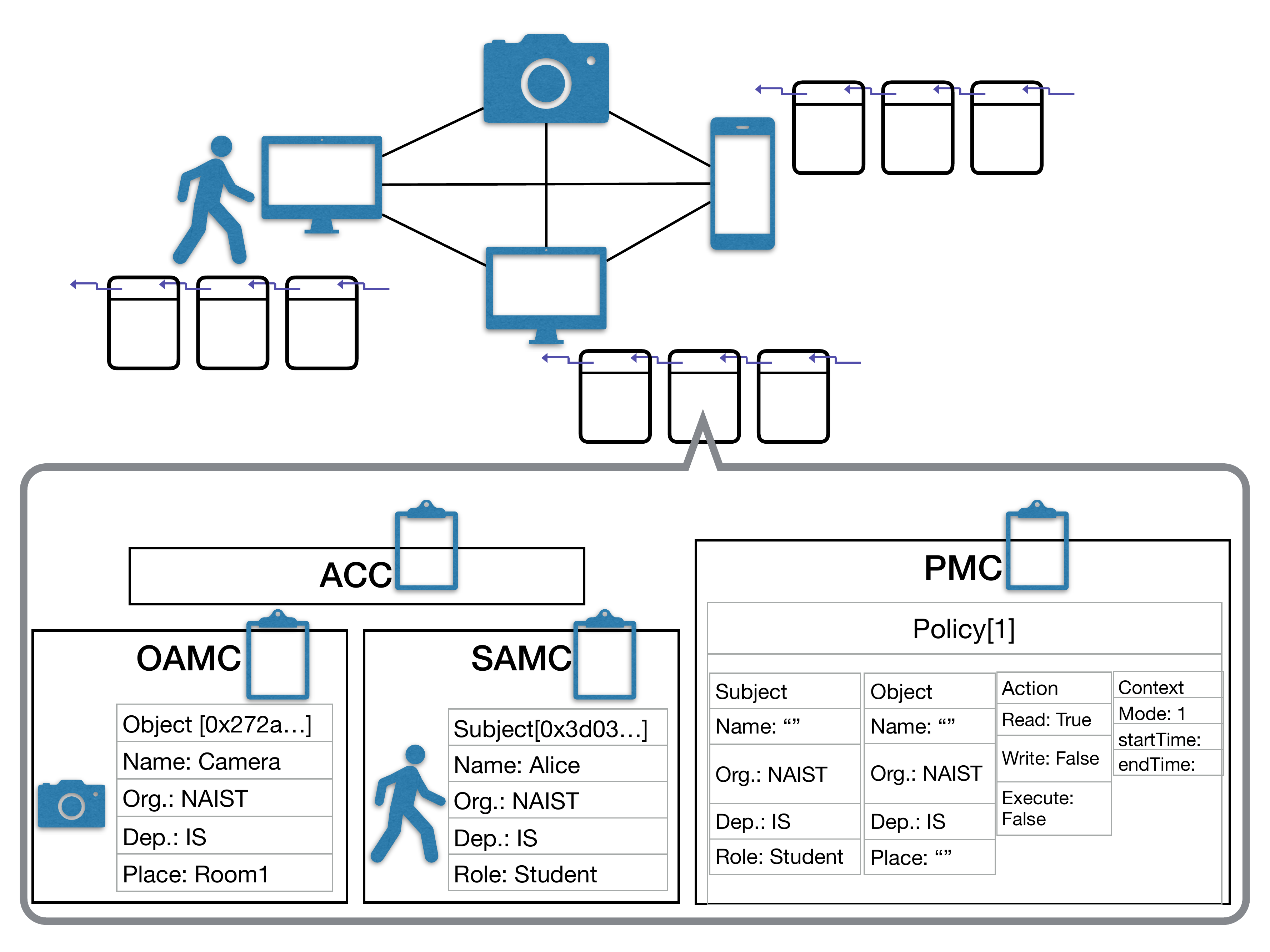}
 \caption{The proposed ABAC framework.}
 \label{fig:propose}
\end{figure}

Fig. \ref{fig:propose} illustrates the proposed ABAC framework, which is composed of four smart contracts, i.e., SAMC, OAMC, PMC and ACC.  
The SAMC, OAMC and PMC are used to store and manage (e.g., update, add and delete) the attributes of subjects, the attributes of objects and the ABAC policies, respectively. 
The ACC performs the access control by interacting with the other three smart contracts. 
This paper considers a smart campus scenario as part of a smart city. 
Note that the application of the framework is not limited to smart campus but can be extended to the whole smart city by properly defining the policies and attributes. 
We introduce the details of the four smart contracts in what follows.

%\begin{figure}[!t]
%\centering
%\includegraphics[width=3in]{./figures/system.eps}
%\caption{Illustration of the considered IoT system.}
%\label{fig_sysmodel}
%\end{figure}

\subsubsection{Subject Attribute Management Contract (SAMC)}

The  SAMC is responsible for the storage and management of the subject attributes in the \yet{smart campus}. 
The SAMC is usually deployed by the administrators of subjects on the blockchain and only the administrators have permissions to execute it. 
In the case of a smart \yet{campus}, the administrators can be the executive office if the subjects are students and staffs, while if the subjects are IoT devices, the administrators can be their owners. 
To distinguish between different subjects in the system, each of them is assigned a unique identifier (i.e., ID). 
In addition, each ID is linked with multiple attributes to indicate that the subject with that ID possesses those attributes.
We use Ethereum account addresses (e.g., the 0x3d03... in Table \ref{tb:attribute}) as such ID information throughout this paper. 
Table \ref{tb:attribute} shows some examples of the subject attributes, such as the organization \textit{Org.} (e.g., Nara Institute of Science and Technology: NAIST), department \textit{Dep.} (e.g., Information Science: IS) and laboratory \textit{Lab.} (e.g., Large-scale Systems Management: LSM) to which the subject belongs as well as the \textit{Role} (e.g., student and staff) of the subject.
In addition to the storage of the subject attributes,  the ABIs of \textit{subjectAdd()} and \textit{subjectDelete()} are also provided to add/update and delete the subject attributes, respectively.

\begin{table}[t]
 \caption{Examples of subject and object attributes}\label{tb:attribute}
 \begin{center}
   \begin{tabular}{|l|l|} \hline
  SubjectList[0x3d03...] & ObjectList[0x272a...] \\ \hline
  Name:``Alice'' & Name: ``Camera''\\
  Org.: ``NAIST'' & Org. ``NAIST'' \\
  Dep.: ``IS'' & Dep.: ``IS'' \\
  Lab.: ``LSM'' & Lab.: ``LSM'' \\
  Role: ``student'' & Place: ``Room1'' \\
  Others: ``'' & Others: ``'' \\ \hline
 \end{tabular}
 \label{samc}
 \end{center}
\end{table}

\subsubsection{Object Attribute Management Contract (OAMC)}
Similar to the SAMC, the role of the OAMC is storing and managing the attributes of the objects, which can be executed only by the object administrators.
Each object also has multiple attributes linked with its ID, i.e., an Ethereum account address (e.g.,  the 0x272a... in Table \ref{tb:attribute}) in this paper. 
This paper considers some examples of the object attributes in Table \ref{tb:attribute}, including the organization \textit{Org.} (e.g., NAIST), department \textit{Dep.} (e.g., IS) and laboratory \textit{Lab.} (e.g., LSM) to which the object belongs as well as the \textit{Place} (e.g., Room1) where the object is placed. 
In addition to the storage of the object attributes, the ABIs of \textit{objectAdd()} and \textit{objectDelete()} are also provided to add/update and delete the object attributes, respectively.

\subsubsection{Policy Management Contract (PMC)}
The PMC is responsible for the storage and management of the ABAC policies defined in this paper.
Similar to the SAMC and OAMC, only the administrators (e.g., the object owners) of the policies have permissions to execute the PMC. 
A policy is a combination of a set $SA$ of subject attributes, a set $OA$ of object attributes, a set $A$ of actions and a set $C$ of context information.
This combination states that the subjects with attributes in $SA$ can perform the actions in $A$ on the objects with attributes in $OA$ under the context in $C$.  
For simplicity, we adopt time as the simple context information for dynamic access control in this paper. 
We use three parameters to represent time, which are \emph{Mode}, \emph{startTime} and \emph{endTime}. 
The \emph{Mode} parameter indicates whether dynamic access control is used. If the \emph{Mode} is set to $0$, dynamic access control is not applied. 
If the \emph{Mode} is set to $1$, dynamic access control is applied and the parameters \emph{startTime} and \emph{endTime} need to be further specified to indicate the start time and end time of the allowed access session. 
That is, access is allowed only if it is during the period between the \emph{startTime} and \emph{endTime}.

Table \ref{tb:policy} shows an example of the ABAC policy defined in our framework with \textit{$SA=\{$Org.: NAIST, Dep.: IS, Lab.: LSM, Role: Student$\}$}, \emph{$OA=\{$Org.: NAIST, Dep.: IS, Lab.: LSM$\}$}, \textit{$A=\{$Read, Write$\}$} and \textit{$C=\{$Mode: 1, startTime: 1563206776, endTime: 1575483330$\}$}. 
Note that the \emph{startTime} and \emph{endTime} are expressed in unixtime format in this example, which correspond to $5/24/2019\ 12:00$ and $5/31/2019\ 11:59$, respectively.
The policy states that any \textit{student} belonging to the \textit{LSM} laboratory of the \textit{IS} department of the \textit{NAIST} organization can \textit{read} and \textit{write} any object at any place of the same laboratory of the same department and organization for a specified period of time between $5/24/2019\ 12:00$ and $5/31/2019\ 11:59$.
Unlike the policies in \cite{wang2019anattribute}, the policies in our framework are not associated with certain subjects or objects. Thus, each policy can handle the access control between multiple objects and multiple subjects, achieving many-to-many access control. As a result, policy search is required for the processing of access requests from subjects.

The \yet{policies} are stored as a list structure in the PMC and managed by the ABIs of \textit{policyAdd()}, \textit{policyDelete()} and \textit{policyUpdate()}, respectively. 
In addition, the ABIs of \textit{findExactMatchPolicy()} and \textit{findMatchPolicy()} are also provided for searching policies. 
We will introduce these two types of policy search in Section \ref{sec:find} in greater details.

\begin{table}[t]
 \centering
 \caption{Example of an ABAC policy.}
 \label{tb:policy}
%\scalebox{0.85}[0.9]{
 \begin{tabular}{|l|l|l|l|}
  \hline
  Subject Attributes  & Object Attributes & Action & Context \\ \hline
  Name: ``'' & Name: ``'' & Read: True & Mode: 1 \\
  Org.: ``NAIST'' & Org.: ``NAIST'' & Write: True & startTime: \\
  Dep.: ``IS'' & Dep.: ``IS'' & Execute: False & 1563206776 \\
  Lab.: ``LSM'' & Lab.: ``LSM'' & & endTime: \\
  Role: ``Student'' & Place: ``'' & & 1575483330\\ \hline
 \end{tabular}
 %}
\end{table}

\subsubsection{Access Control Contract (ACC)}
The core of the access control system is the ACC, which is responsible for controlling the access requests from the subjects to the objects. 
To execute the ACC, a transaction that contains the required request information (e.g., subject ID, object ID and actions) must be sent to the \textit{accessControl()} ABI. 
When the ACC receives the transaction, it will retrieve the corresponding subject attributes, object attributes and the policy from the SAMC, OAMC and PMC, respectively. 
Based on the attributes and policy, the ACC decides if the subject is allowed to perform the requested actions on the object. 
Finally, such decision results will be returned to both the subject and object for reference. 
We will introduce the access control flow in Section \ref{sec:ac} in greater details.

\subsection{Main Functions of the Framework}
\label{sec:main}
The proposed ABAC framework provides the following main functions.

\subsubsection{Adding, Updating and Deleting Subject/Object Attributes}
As mentioned in Section \ref{sec:sc_system}, the basic functions provided by the proposed ABAC framework are adding, updating and deleting the attributes of the subjects/objects. 
For example, to add/update the attributes of a subject, the subject administrator can send a transaction, which contains the subject's ID and the attributes to add/update, to the \textit{subjectAdd()} ABI of the SAMC. 
If the ABI finds a matched entry for the presented subject ID in the subject list, it will update the attributes of the subject.
Otherwise, the ABI will create a new entry for the subject ID in the subject list.
When deleting some attributes of a subject, the subject administrator can send another transaction, which contains the subject's ID and attributes to delete, to the \textit{subjectDelete()} ABI.
Examples of adding, updating and deleting the attributes of an object are quite similar to the above ones and are thus omitted here.

\subsubsection{Searching Policies}
\label{sec:find}
Because the PMC stores the policies as a list (i.e., array), we need policy search to delete, update and retrieve a certain policy. 
The proposed ABAC framework provides the ABIs of \textit{findExactMatchPolicy()} and \textit{findMatchPolicy()} to implement two patterns of policy search, respectively, i.e., search by complete match and search by partial match.

\begin{itemize}
\item Search by Complete Match:
This policy search pattern returns the policies with subject and object attributes exactly matching the subject and object attributes provided by the transaction sender. 
For example, when a transaction sender searches for the policy as illustrated in Table \ref{tb:policy}, he/she needs to provide exactly the same attribute information as listed in Table \ref{tb:policy}. 
This search pattern is mainly used for deleting policies.

\item Search by Partial Match:
Differing from the previous policy search pattern, this pattern returns a list of indices of the policies, in which the subject and object attributes are subsets of those provided by the transaction sender.  
Suppose a transaction sender executes this search by offering a set $SA$ of subject attributes and a set $OA$ of object attributes. 
Taking $SA$ and $OA$ as inputs, the search will return any policy whose subject attribute set $SA'$ and object attribute set $OA'$ satisfy the \yet{conditions} $SA'\subseteq SA$ and $OA'\subseteq OA$. 
The reason is that any found policy can handle the access request from the subjects with attribute set $SA$ to the objects with attribute set $OA$. 
This search pattern is mainly used for adding/updating policies and access control by the ACC. 
\end{itemize}

\subsubsection{Adding, Updating and Deleting Policies}
A policy is defined as a logical combination of subject attributes, object attributes, actions and contexts. 
In general, policy administrators can define more fine-grained policies by including more attributes, thus achieving more flexible and dynamic access control. 
Similar to the attribute management, this framework also provides functions for policy management, including adding, updating and deleting policies.
When adding a new policy, the policy administrator first needs to execute the policy search by partial match (i.e.,  \textit{findMatchPolicy()}) to find the similar policies. 
When any similar policies are returned, the administrator then needs to ensure that the new policy to add does not conflict with any of the returned similar policies. 
Any possible conflicts must be resolved by the administrator.
After the conflict resolution, the policy is finally added to the policy list by the administrator through the \textit{policyAdd()} ABI of the PMC. 

Similarly, to update a policy, the administrator also needs to apply the search by partial match to find the target policy.
Note that other similar policies will also be returned by the search as well. 
Conflicts (if any) between the new policy used for update and the similar policies must be resolved, after which the target policy is then replaced by the new one through the \textit{policyUpdate()} ABI.
Note that when existing policies cover the new ones to add/update, adding/updating policies is not required, which can reduce the monetary cost and the storage overhead of the framework. 
Different from adding/updating policies, the policy administrator needs to apply the policy search by complete match to find the target policy, when deleting a policy. 
If the target policy is found, the administrator then deletes it from the policy list by executing the \textit{policyDelete()} ABI.

\subsubsection{Access Control}
\label{sec:ac}
Access control is the core function of the proposed ABAC framework. 
To illustrate the typical access control flow, we show in Fig. \ref{fig:model} a subject \textit{Alice} with attributes \textit{Role: Student}, \textit{Dep: IS} and \textit{Org: NAIST} (i.e., a student belonging to the \textit{IS} department of \textit{NAIST}) wishes to access an object camera with attributes \textit{Place: Room1}, \textit{Dep: IS} and \textit{Org: NAIST} (i.e., an object located in \textit{Room} $1$ of the \textit{IS} department of \textit{NAIST}). 
We introduce the steps of the flow in greater details as follows:

\begin{figure}[t]
 \centering
 \includegraphics[width=\columnwidth]{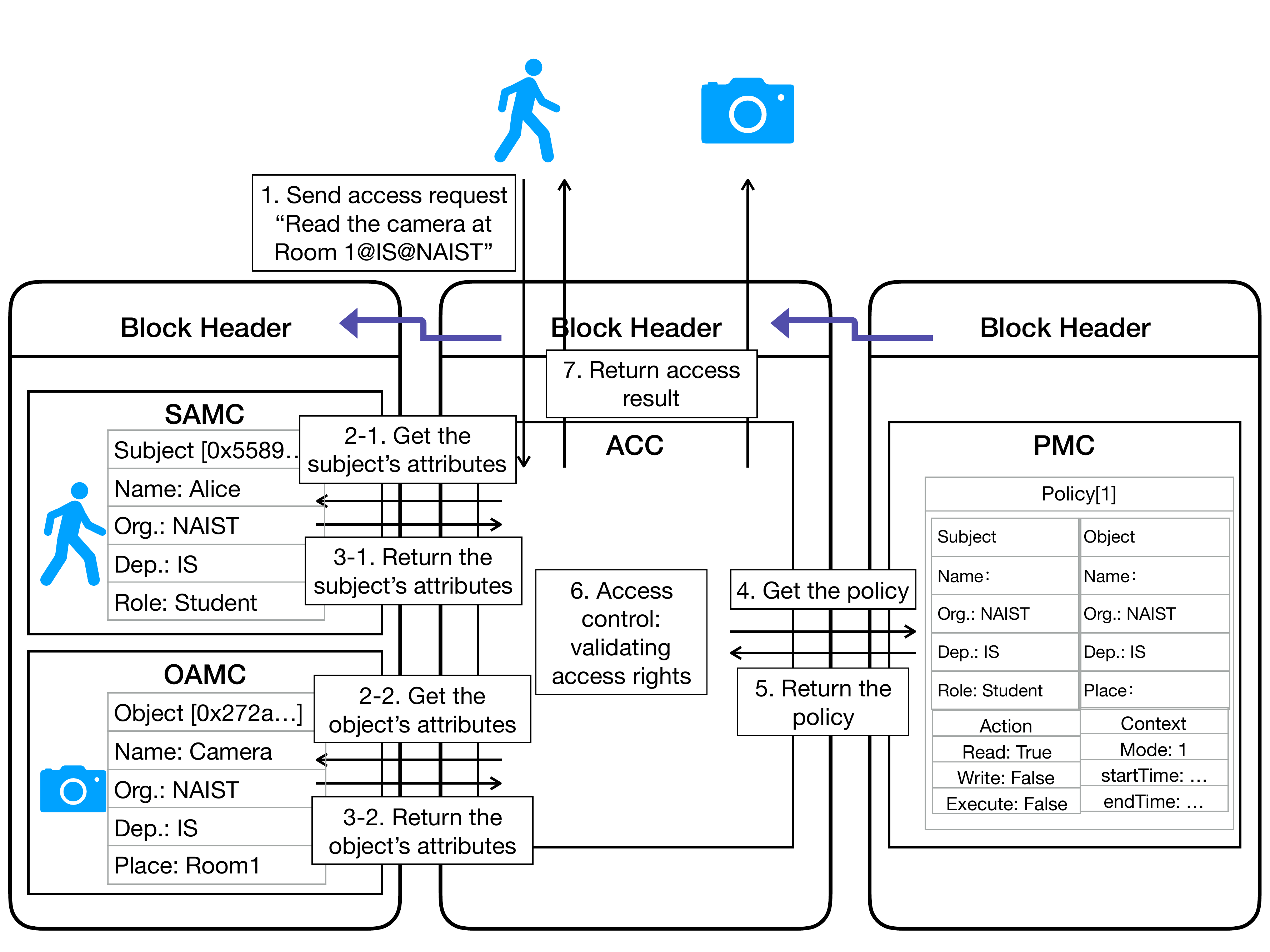}
 \caption{Access control process.}
 \label{fig:model}
\end{figure}

\begin{itemize}
   \item Step 1: Alice sends an access request to the ACC by sending a transaction, which contains her ID, the object ID and the actions to perform, to the \textit{accessControl()} ABI.
   \item Step 2: The ACC retrieves the subject (resp. object) attributes from the SAMC (resp. OAMC) by sending a message containing the subject (resp. object) ID to the \textit{getSubject()} (resp. \textit{getObject()}) ABI.
   \item Step 3: The SAMC and OAMC return the corresponding subject attributes and object attributes to the ACC, respectively.
   \item Step 4: The ACC sends a message, which contains the attributes of the subject and object, to the \textit{getPolicy} ABI of the PMC to query the related policies.
   \item Step 5: The PMC applies the policy search by partial match to search the related policies and returns the found policies to the ACC.
   \item Step 6: Based on the returned policies as well as the subject attributes, object attributes and context information (e.g., time), the ACC determines if the subject has rights to perform actions on the object.
   \item Step 7: The ACC returns the access results to both the subject and object.
\end{itemize}

%Using the smart contract system, the access control becomes a distributed application, which is executed by the majority of the system nodes. 
%In addition, the access history and results are also stored on the blockchain. 
%Thus, even if some of the nodes are destroyed by disasters or compromised by adversaries, the access control framework can still work reliably. 
%This achieves distributed and trustworthy access control for the IoT.

\subsubsection{Dynamic Access Control}
This framework implements simple dynamic access control based on the time-related variables of Solidity. 
By checking the \emph{Mode} parameter in the context field of the policy, the ACC determines whether dynamic access control will be executed. 
If the mode is $0$, dynamic access control is not required and the ACC performs the access control solely based on the action field of the policy. 
If the mode is set to $1$, the ACC then performs dynamic access control based on the fields of action and context. 
More specifically, the ACC obtains the current time using the \emph{now} variable provided by the Solidity and check if the inequality $startTime \leq now \leq endTime$ holds. 
If the answer is yes, the request passes the dynamic control. 
Otherwise, the request will be denied.

\section{Implementation}\label{sec:implementation}

\begin{figure}[t]
 \centering
 \includegraphics[width=0.6\columnwidth]{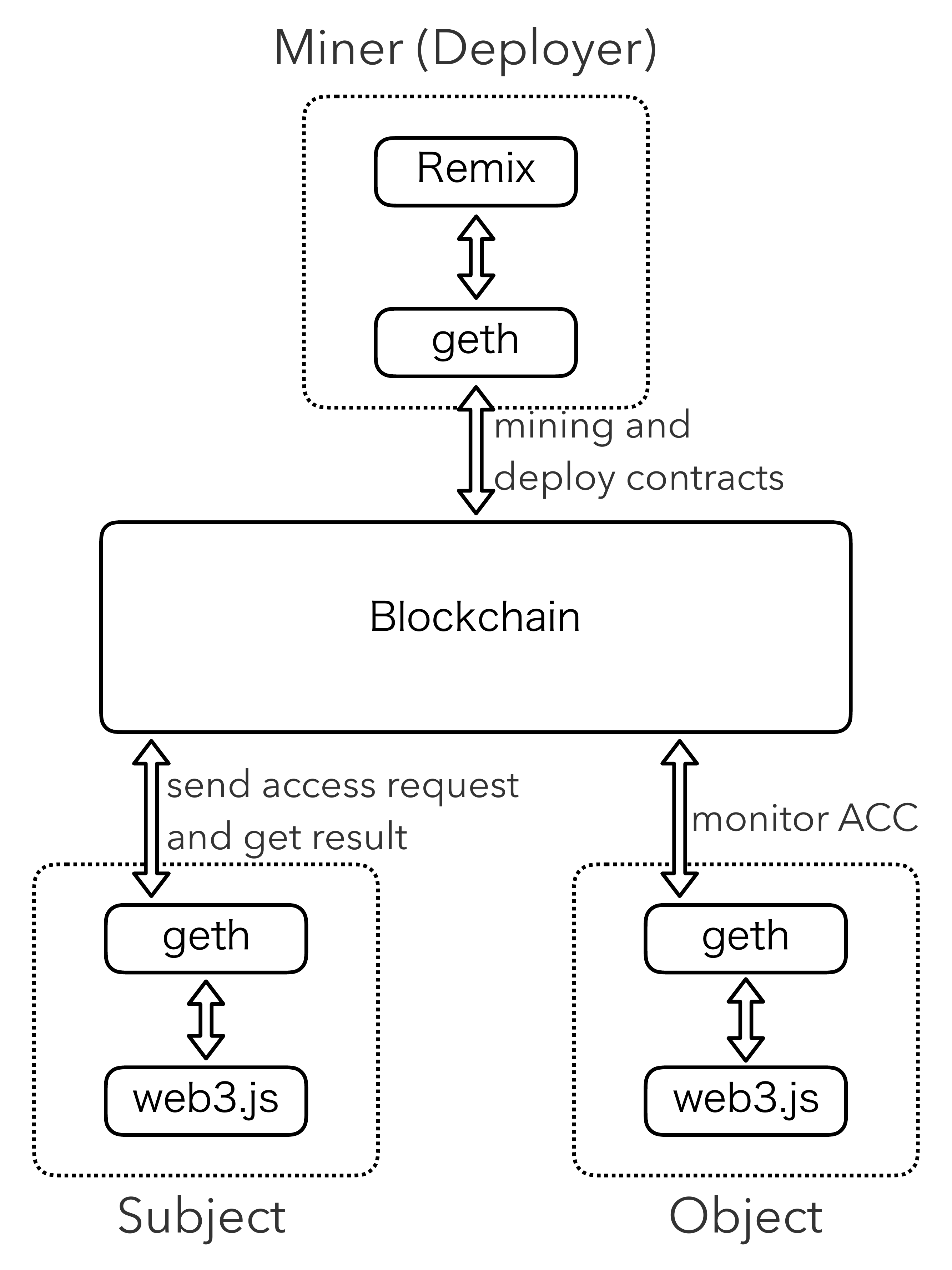}
 \caption{Software used in the proposed framework.}
 \label{fig:software}
\end{figure}

To implement the proposed ABAC framework, we used the geth client \cite{geth}  to set up three Ethereum nodes on a local server (Intel Xeon CPU E5-1620 3.60 GHz, 32 GB memory), such that  a private Ethereum blockchain network can be constructed as shown in Fig.  \ref{fig:software}. 
One node plays the role of the miner of the blockchain network and the other two serve as the subject and object, respectively. 
We used the the Remix \cite{Remix}, a browser-based Integrated Development Environment (IDE), to edit and compile the four smart contracts of the proposed framework.
Since the Remix IDE can be configured to connect to any of the three nodes via a Remote Procedure Call (RPC) connection, we used the Remix IDE to deploy the four smart contracts on the private blockchain.
To interact with the nodes, we also created JavaScript programs using the web3.js package \cite{web3}, which can be used for sending transactions and viewing the access result at both the subject and object sides.

\begin{figure}[t]
 \centering
 \includegraphics[width=\columnwidth]{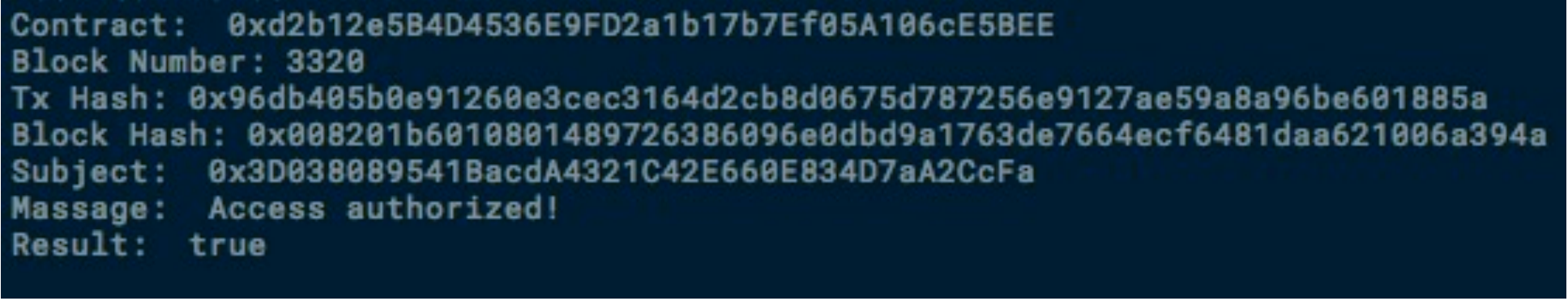}
 \caption{Result of valid access (action: read).}
 \label{fig:true}
\end{figure}

\begin{figure}[t]
 \centering
 \includegraphics[width=\columnwidth]{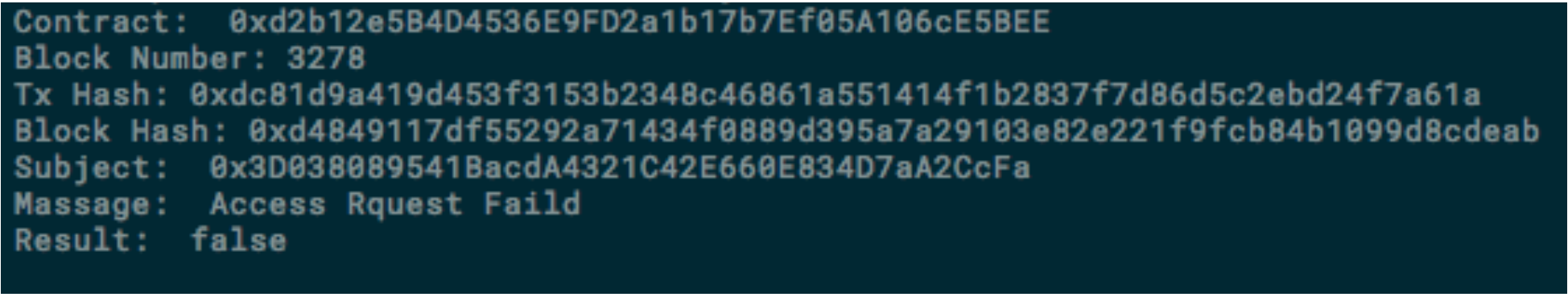}
 \caption{Result of invalid access (action: execute).}
 \label{fig:false}
\end{figure}

Based on the deployment, we conducted experiments to demonstrate the feasibility of the proposed framework.
During the experiments, we adopted the attributes in Table \ref{tb:attribute} as the subject and object attributes and also used the policy in Table \ref{tb:policy} for the access control.
We can see from Tables \ref{tb:attribute} and \ref{tb:policy} that the subject can only perform the \textit{read} and \textit{write} actions on the object. 
We show in  Fig. \ref{fig:true} (resp. Fig. \ref{fig:false}) the results of the case where the subject sends a \textit{read} (resp. \textit{execute}) request. 
The outputs in both figures contain various information, including the address of the ACC, the hash of the access request transaction sent by the subject, the hash of the block where the access request transaction is stored, the address (i.e., ID) of the subject and the access results. 
The results in these two figures demonstrate the feasibility of the proposed ABAC framework.

\section{Cost Evaluation}\label{sec:evaluation}
The system administrators need to pay some money (also known as transaction fee) to deploy smart contracts on the blockchain and execute the ABIs of these contracts. 
Ethereum uses a unit called \textit{gas} to measure the amount of operations needed to perform a task, e.g., deploying a smart contract or executing an ABI. 
In general, the more complex a task is, the more gas it consumes. 
Like the gas at gasoline stations, gas in Ethereum also has price, which varies with time. 
A higher gas price usually makes transactions mined faster. 
Thus, as shown in the following expression, the money required  for performing a task, denoted by $TxFee$, is the product of the amount of consumed gas, denoted by $gas$, and the gas price, denoted by $gasPrice$.
\begin{IEEEeqnarray}{rCl}\label{eqn:txfee-calculation}
TxFee = gas \times gasPrice \times 10^{-9}.
\end{IEEEeqnarray}
Note that the $TxFee$ has the unit $Ether$, and $gasPrice$ has the unit $Gwei$. Thus, the term $10^{-9}$ in the above equation is needed for unit conversion. Unless otherwise stated, we set the $gasPrice$ to $8$ $Gwei$.

In general, the gas cost comes from three aspects, which are code cost denoted by $codeCost$, storage cost denoted by $storageCost$ and initial cost denoted by $initCost$. 
The code cost is related to the complexity of the code executed by the transaction. 
The more complex the code is, the higher the code cost will be. 
The storage cost is the gas consumption for modifying the storage of the smart contract, such as addition, deletion and modification of data. 
More storage modification incurs higher storage cost. 
Finally, the initial cost represents the gas consumed when some ABIs are executed for the first time. 
Thus, the gas consumption can be expressed as
\begin{IEEEeqnarray}{rCl} \label{eqn:gas_cons}
gas = codeCost + storageCost + initCost.
\end{IEEEeqnarray}

In this section, we will show the cost required to execute each operation according to the flow of Section \ref{sec:main}, and compare the cost required for access control of this framework and that of the ACL-based framework in \cite{zhang2018smart}, including the deployment cost for deploying smart contracts and operating cost during the running of the access control frameworks. 
We first estimate the expression of the gas consumption of each operation based on (\ref{eqn:gas_cons}) and then obtain the parameters inside the expressions from experimental results.
%and user cost incurred when users use the frameworks. 
Table \ref{tb:pmts} summarizes the parameters used in this section and their meanings.

\begin{table}[t]
 \centering
  \caption{Parameters and their meanings}
 \label{tb:pmts}
 \begin{tabular}{rl} \hline
  variable & definition \\ \hline \hline
  $A_s$ & Maximum number of subject attributes \\
  $A_o$ & Maximum number of object attributes  \\
  $C_s$ & Maximum number of characters in one subject attribute\\
  $C_o$ & Maximum number of characters in one object attribute \\
  $l$ & Length of the policy list \\
  $n$ & The number of policies to add \\
  $m$ & the number of subject-object pairs \\ \hline
 \end{tabular}
\end{table}

\subsection{Cost of Each ABI}
\subsubsection{Adding, Updating and Delete Attributes}
\begin{itemize}
\item Adding the attributes of a subject: The \textit{subjectAdd()} ABI is executed to add the attributes of a subject to the \emph{subjectList}. 
Because the \textit{subjectAdd()} ABI will modify the storage (i.e., the \emph{subjectList}) of the smart contract, the gas cost depends on the number of attributes as well as the number of characters in an attribute. 
Denoting the maximum number of attributes of a subject by $A_s$ and the maximum number of characters allowed in one subject attribute by $C_s$, the upper bound on the gas cost of adding subject attributes can be expressed as
\begin{IEEEeqnarray}{rCl}
G_{SA} &= 64 \times A_s\times C_s + SA(A_s), 
\end{IEEEeqnarray}
where $SA(A_s)$ represents the $codeCost$ depending on $A_s$, $64$ represents the cost for adding one character and $64 \times A_s\times C_s$ represents the $storageCost$. 
In our experiment, we set $A_s=6$, which results in $SA(A_s)=151,250$.

\item Updating a subject attribute: The \textit{subjectUpdate()} ABI is executed to rewrite a subject attribute in the \textit{subjectList}. 
The gas cost of this ABI also depends on the number of characters in the newly input attribute, which can be expressed as
\begin{IEEEeqnarray}{rCl}
G_{SU} &= 61,250 + 64 \times C_s,
\end{IEEEeqnarray}
where $61,196$ represents the $codeCost$ and $64 \times C_s$ represents the $storageCost$.

\item Deleting a subject attribute: The \textit{subjectDelete()} ABI is executed to delete an attribute of a subject from the \textit{subjectList}, which depends only on the $codeCost$ and thus consumes a constant amount of Gas, about $26,786$ in our experiments.

\item Adding the attributes of an object: The \textit{objectAdd()} ABI is executed to add the attributes of an object to the \textit{objectList} on the blockchain. 
Similar to the \textit{subjectAdd()} ABI, the \textit{objectAdd()} ABI will modify the storage (i.e., the \textit{objectList}) of the smart contract, and thus the gas cost depends on the number of attributes as well as the lengths of the attributes to add. 
Denoting the maximum number of attributes of an object by $A_o$ and the maximum number of characters allowed in one object attribute by $C_o$, we can obtain the upper bound on the gas cost of adding an object attribute as
\begin{IEEEeqnarray}{rCl}
G_{OA} &= 64 \times A_o\times C_o + OA(A_o),
\end{IEEEeqnarray}
where $OA(A_o)$ represents the $codeCost$ depending on $A_o$ and $64 \times C_o$ represents the $storageCost$. 
In our experiment, we set $A_o=6$, which results in $OA(A_o)=151,228$.

\item Updating an object attribute: The \textit{objectUpdate()} ABI is executed to rewrite the object attribute in the \textit{objectList}. 
Similar to the \textit{subjectUpdate()} ABI, the gas cost of the \textit{objectUpdate()} ABI depends on the number of characters in the newly input attribute, which is upper bounded by
\begin{IEEEeqnarray}{rCl}
G_{OU} &= 61,228 + 64 \times C_o,
\end{IEEEeqnarray}
where $61,228$ represents the $codeCost$ and $64 \times C_o$ represents the $storageCost$.

\item Deleting an object attribute: The \textit{objectDelete()} ABI is executed to delete an attribute of an object from the \textit{objectList}, which, similar to the \textit{subjectDelete()} ABI, consumes a constant amount of $codeCost$, about $26,808$ in our experiments.

\end{itemize}

\subsubsection{Adding, Updating and Deleting Policies}
\begin{itemize}
\item Adding a policy: The \textit{policyAdd()} ABI is executed to add a new policy to the \textit{policyList}. 
Some initial processing is required in this ABI, so it consumes some initial cost when it is executed for the first time. 
In addition, this ABI also uses some storage to store the policy. 
Thus, it also consumes some storage cost, which varies with the numbers of subject and object attributes as well as the total number of characters in the policy. 
The following equation expresses the upper bound on the gas cost of adding a policy:
\begin{IEEEeqnarray}{rCl}
G_{PA} &=& 213,803 + 15,000 \times (A_s + A_o) \\
&&+ 64 \times ( A_s \times C_s + A_o \times C_o) \nonumber\\
&& + 15,000 \times (A_s + A_o + 1) \times \mathbf{1}_{first time},\nonumber
\end{IEEEeqnarray}
where $213,803$ represents the $codeCost$, $15,000 \times (A_s + A_o) + 64 \times ( A_s \times C_s + A_o \times C_o) $ represents the $storageCost$ and $\mathbf{1}_{first time}$ is the indicator function, which equals $1$ when the \textit{policyAdd()} ABI is executed for the first time.

\item Updating a policy:
The \textit{policyUpdate()} ABI is executed to rewrite the policy in the \textit{policyList}, where the gas cost depends on the number of characters of the newly input policy. 
The upper bound on the gas cost of updating a policy can be expressed as
\begin{IEEEeqnarray}{rCl}
G_{PU} = \left \{
\begin{array}{l}
 194,337 + 64 \times \\( A_s \!\times \!C_s \!+\! A_o \!\times\! C_o), \, policy's\, index=0,\\
 194,401 + 64 \times \\( A_s \!\times \!C_s \!+\! A_o \!\times\! C_o),  \, policy's\, index\ge1,
 \end{array}
 \right.
\end{IEEEeqnarray}
where $64 \times ( A_s \times C_s + A_o \times C_o)$ represents the $storageCost$, $194,337$ and $194,401$ represent the $codeCost$, which varies depending on the index of  the policy to update.

\item Deleting a policy: The \textit{policyDelete()} ABI is executed to delete a policy from the \textit{policyList}, which consumes a constant cost, as expressed in the following equation.
\begin{IEEEeqnarray}{rCl}
G_{PD} = \left \{
\begin{array}{l}
 51,529,\, policy's\,index=0, \\
 51,561, \, policy's\,index\ge1.
 \end{array}
 \right.
\end{IEEEeqnarray}
Note that the gas cost for deleting a policy varies depending on the index of the policy to delete.
\end{itemize}

\subsubsection{Searching Policies}
\begin{itemize}
\item Searching a policy: The \textit{findPolicy()} is the ABI for searching the corresponding policy based on the attributes of the subjects and objects required by the ACC. 
The gas cost depends on the length and the contents of the policy to be searched. 
Denoting the length of \textit{policyList} by $l$, the upper bound on the gas cost of searching a policy can be expressed as
\begin{IEEEeqnarray}{rCl}
G_{FP}(l) &=& 57,495+ 4,000\times (A_s + A_o) \\
&&+ 10,518\times l + 64\times ( A_s \times C_s + A_o \times C_o),\nonumber
\end{IEEEeqnarray}
where $57,495$ represents the $codeGas$, $4,000$ represents the unit cost for searching with one subject attribute or object attribute, $10,518$ represents the unit cost for searching in a policy list of length one and $64$ represents the cost for searching one character. 
The $A_s \times C_s + A_o \times C_o$ here denotes the maximum number of characters in the policy.

\item Adding multiple policies: Before adding a policy, a search in the policy list is required to find if the policy has been added. 
Thus, the gas cost for adding policies depend also on the cost of searching policies, i.e., the $G_{FP}(l)$. 
When $n$ distinctive policies need to be added, the upper bound on the gas cost can be expressed as
\begin{IEEEeqnarray}{rCl}
\lefteqn{ \nonumber n \times G_{PA} + \sum_{k=1}^{n} G_{FP}(k) } \\
\nonumber&=& \sum_{k=1}^{n} G_{FP}(k) + n \times \Big(213,803 + 15,000 \times (A_s + A_o)\\
\nonumber && \ \ \ \ \ \ \ \ \ \ \ \ \ \ \ \ \ \ \ \ \ + 64 \times ( A_s \times C_s + A_o \times C_o) \\
\nonumber &&  \ \ \ \ \ \ \ \ \ \ \ \ \ \ + 15,000 \times (A_s + A_o + 1) \times \mathbf{1}_{first time}\Big) \\
&=& 5,259n^2 \! +\! \Big( 276,557 \!+ \!19,000 \!\times\! (A_s \!+\! A_o) \\
\nonumber&&\ \ \ \ \ \ \ \ \ \ \ \ \ \ \ +  128 \!\times \!(A_s \!\times\! C_s \!+\! A_o \!\times\! C_o) \\
\nonumber &&\ \ \ \ \ \ \ \  + 15,000 \times (A_s + A_o + 1)\times \mathbf{1}_{first time}\Big)n.
\end{IEEEeqnarray}

\end{itemize}
\subsubsection{Access control}
In this access control framework, it is necessary to execute five ABIs to conduct access control, i.e., \textit{getSubject()}, \textit{getObject()}, \textit{findPolicy()}, \textit{getPolicy()} and \textit{accessControl()}. 
The cost required to execute each ABI is as follows:
\begin{itemize}
\item \textit{getSubject()}:
When the ACC wants to  obtain subject attribute information from the SAMC, it executes the \textit{getSubject()} ABI. 
When this ABI is executed for the first time, it is necessary to create a variable $subject$ in the ACC. 
Thus, an $initCost$ of $15,000$ is required as shown in the following equation, which expresses the upper bound on the gas cost of the \textit{getSubject()} ABI.
\begin{IEEEeqnarray}{rCl}
G_{GS} = 59,467 + 15,000 \times A_s \times \mathbf{1}_{first\,time},
\end{IEEEeqnarray}
where $59,467$ represents the $codeCost$ and $\mathbf{1}_{first\,time}$ is an indicator function, which equals $1$ if the \textit{getSubject()} ABI is executed for the first time and equals $0$ otherwise.

\item \textit{getObject()}:
The ACC executes the \textit{getObject()} ABI to obtain the object attribute information from the OAMC. 
Like the \textit{getSubject()} ABI, the \textit{getObject()} ABI needs to create a variable $object$ in the ACC when it is executed for the first time, so it costs an $initCost$ of $15,000$. 
The upper bound on the gas cost of the  \textit{getObject()} ABI is
\begin{IEEEeqnarray}{rCl}
G_{GO} = 59,201 + 15,000 \times A_o \times \mathbf{1}_{first time},
\end{IEEEeqnarray}
where $59,201$ represents the $codeCost$.

\item \textit{getPolicy()}:
After searching the corresponding policy with the \textit{findPolicy()} ABI, the ACC executes the \textit{getPolicy()} ABI to get the corresponding policy. The gas cost of this ABI depends on whether the policy is found or not. If the policy is found, $53,215$ gas is consumed. Otherwise, $46,780$ gas will be consumed.

\item \textit{accessControl()}:
After obtaining the policy information, access verification is performed using the \textit{accessControl()} ABI. In access verification, the context information in the policy is compared with the action that the subject wants to execute. Similar to the \textit{getPolicy()} ABI, the required gas varies depending on whether the policy is found or not. If the policy is found, $26,932$ gas is consumed. Otherwise, $26,640$ gas is consumed.
\end{itemize}

\subsection{Deployment Cost}
The deployment cost represents the cost used for deploying the smart contracts, i.e., the ACC, SAMC, OAMC and PMC, when the proposed ABAC scheme is introduced to smart cities. 
\yet{Since each smart contract is deployed independently via a separate transaction, the amount of consumed gas can be obtained directly from the \emph{gasUsed} filed of the output of the deployment transaction.
For comparison, we also evaluated the deployment cost of the ACL-based scheme in \cite{zhang2018smart}. 
The gas of the proposed scheme (resp. the scheme in  \cite{zhang2018smart}) is calculated as the aggregate gas for deploying the ACC, SAMC, OAMC, and PMC (resp. RC and JC).
The deployment gas costs of both schemes are summarized in Table \ref{tb:se1}. 
Also, we calculate the deployment costs in US Dollars (USDs).
To do this, we first apply the equation in (\ref{eqn:txfee-calculation}) to calculate the aggregate transaction fees from the gas costs, which are in units of \emph{Ether}.
We then convert the transaction fees in \emph{Ether} to deployment costs in USD based on the exchange rate between USD and \emph{Ether} as of 17:00 JST, March 31, 2020 \cite{EthGas}.}

\begin{table}[t]
 \centering
 \caption{Deployment cost}
 \label{tb:se1}
 \begin{tabular}{|l|r|r|} \hline
  & Gas & USD\\ \hline
  ACL-based Scheme in \cite{zhang2018smart} & 2,809,093 & 2.9664 \\
  Proposed scheme & 4,943,332 & 5.22016 \\ \hline
 \end{tabular}
\end{table}

Table \ref{tb:se1} shows that the proposed scheme consumes more gas than the scheme in \cite{zhang2018smart} at the initial deployment stage, because more smart contracts need to be deployed in the proposed scheme.

\subsection{Operating Cost}
This subsection evaluates the operating cost of the proposed ABAC framework, i.e., the gas consumed during the operation of the access control system. 
It is obvious that the operating cost depends on the number of subjects and objects of the concerned system. 
Thus, we first evaluate the cost when $m$ pairs of subjects and objects are added to the system at the beginning of the system operation, i.e., right after the initial deployment stage. 
Unless otherwise stated, both the maximum numbers of subject and object attributes are set to $6$, i.e., $A_s=A_o=6$. 
Also, the maximum numbers of characters in a subject attribute and an object attribute are set to $10$, i.e., $C_s=C_o=10$.

\subsubsection{Operating Cost vs. Number of Subject-Object Pairs}
First, the attributes of these subjects and objects need to be added. 
Note that the upper bound on the numbers of subjects and objects are both $m$ for $m$ subject-object pairs. 
Second, the policies corresponding to these pairs need to be added to the policy list. 
Suppose the number of policies required by these pairs is $n$, the upper bound on the cost of adding $m$ pairs of subjects and objects at the beginning of the system operation is expressed as
\begin{IEEEeqnarray}{rCl}
G_{AP} = m \!\times\! ( G_{SA} \!+\! G_{OA} )\!+ \!n \times G_{PA} \!+\! \sum_{k=1}^{n} G_{FP}(k),
\end{IEEEeqnarray}
where $m \times ( G_{SA} +  G_{OA} )$ represents the cost for adding attributes and $n \times G_{PA} + \sum_{k=1}^{n} G_{FP}(k)$ stands for the cost for adding policies, because before adding a policy, it is necessary to perform a policy search to prevent duplicated policies.

In the ACL-based scheme in  \cite{zhang2018smart}, one ACC as well as a policy need to be deployed for each subject-object pair. 
Thus, when $m$ pairs of subject and objects are added to the system, the operating cost is
\begin{IEEEeqnarray}{rCl}
G_{AP}^{ACL} = m \times ( G_{ACC} + G_{P}),
\end{IEEEeqnarray}
where $G_{ACC} = 1,706,290$ represents the cost for deploying one ACC and $G_{P} = 238,777$ represents the cost for adding one policy.

\begin{figure}[t]
 \centering
 \includegraphics[width=\columnwidth]{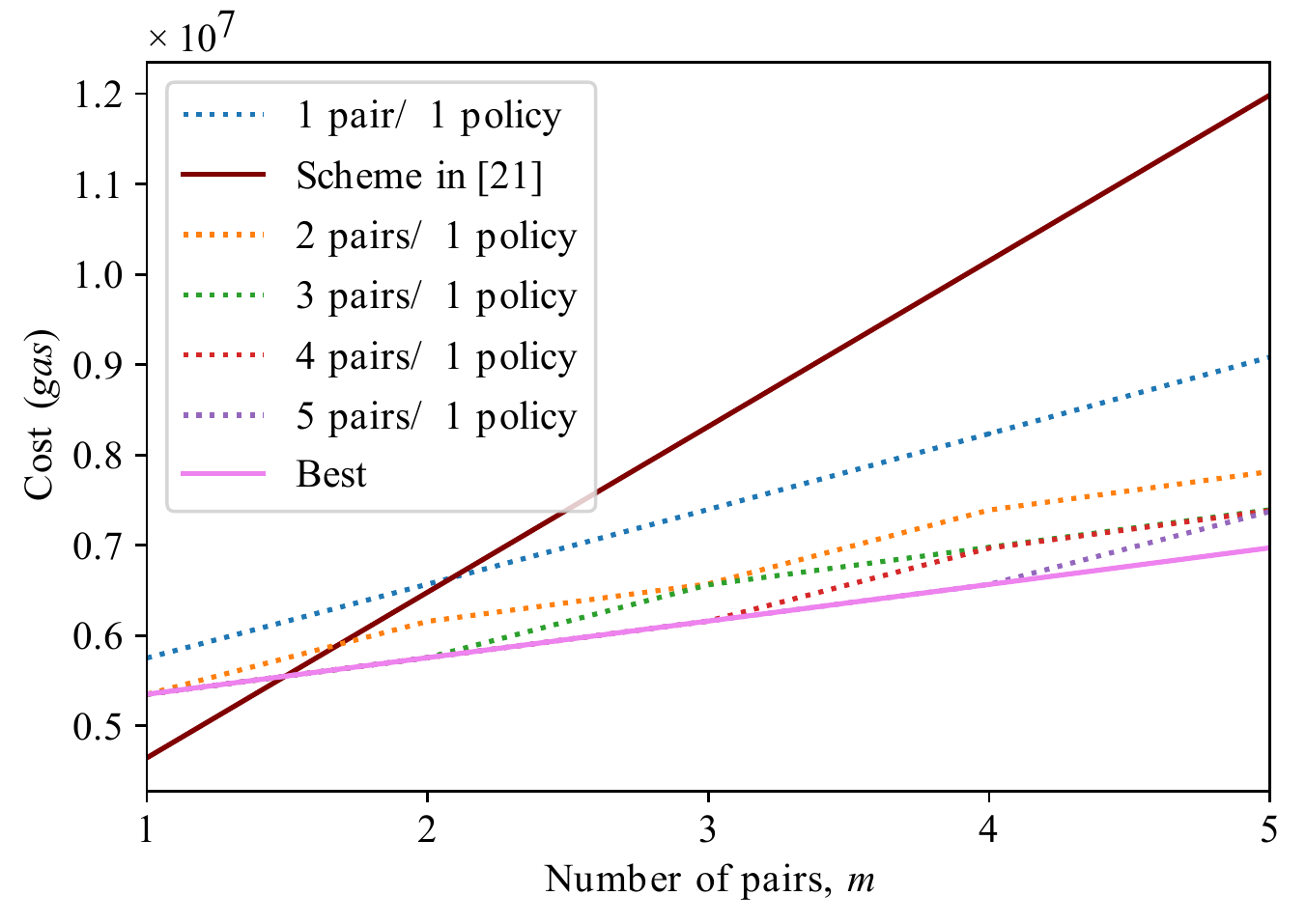}
 \caption{Operating cost vs. number of subject-object pairs, $m$.}
 \label{fig:opeCost_s}
\end{figure}

\begin{figure}[t]
 \centering
 \includegraphics[width=\columnwidth]{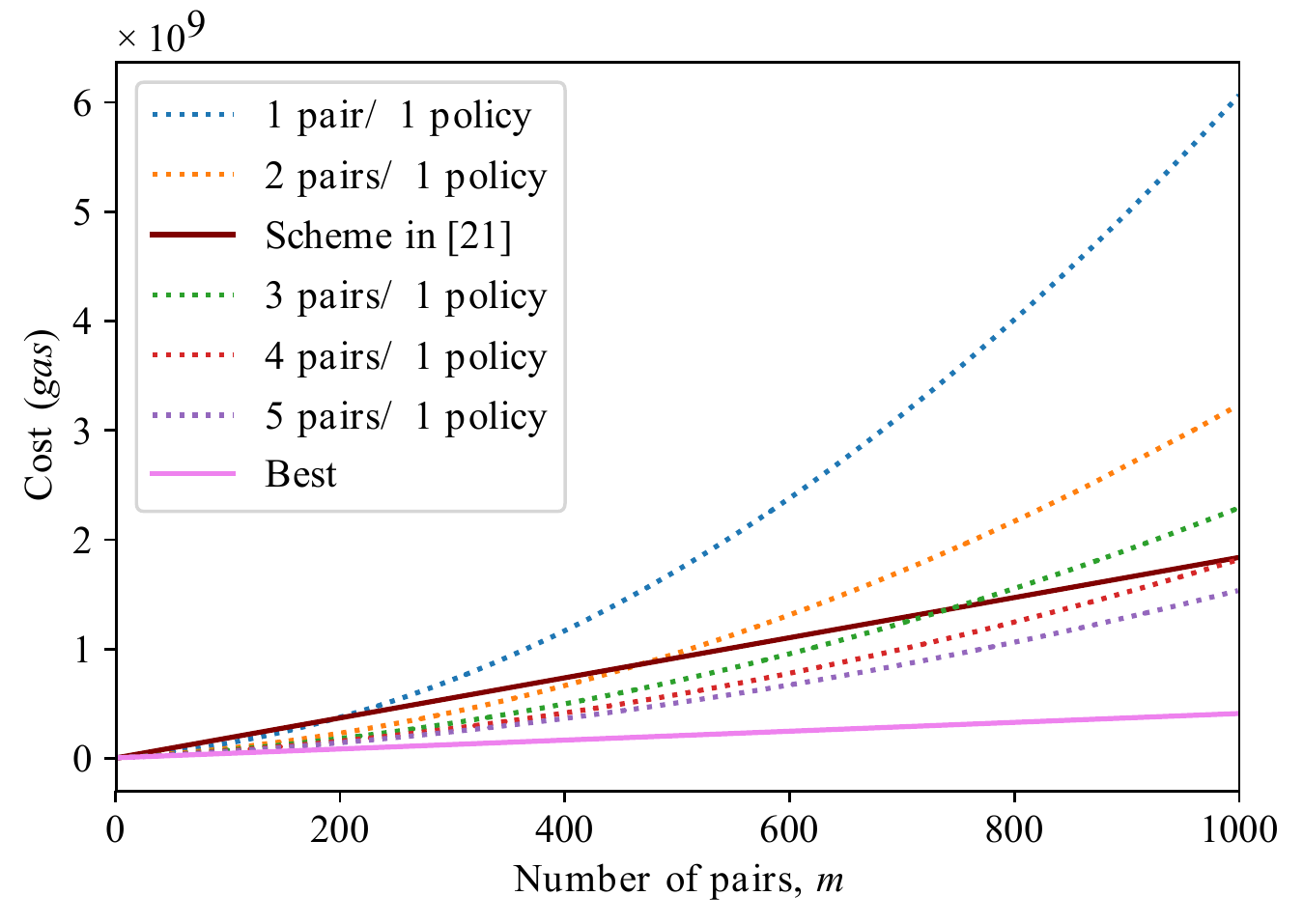}
 \caption{Operating cost with more subject-object pairs.}
 \label{fig:opeCost_l}
\end{figure}

We further evaluate the impacts of the number of subject-object pairs $m$ on the operating cost. 
Fig. \ref{fig:opeCost_s} shows a graph of the administrator's operating cost when $m$ increases. 
We consider six cases for the number of policies $n$ required by $m$ subject-object pairs, i.e.,  $n=m$, $n =m/2$, $n=m/3$, $n=m/4$, $n=m/5$ and $n=m/m$. 
The $n=m/p$ here means that one policy handles the access control of $p$ pairs. 
Note the case of $n=m/m$ is the best case, where all the $m$ pairs share one policy. 
In other words, there is no need to add more policies. 
Fig. \ref{fig:opeCost_s} also shows the cost of the ACL-based scheme in \cite{zhang2018smart}.

When the number of subject-object pairs is less than $3$, the proposed scheme requires deploying four smart contracts, leading to a larger deployment cost, so the administrator's operating cost is larger than that of the ACL-based scheme in \cite{zhang2018smart}. 
However, when the number of subject-object pairs is $3$ or more, the ACC deployment in \cite{zhang2018smart} has a larger impact and thus the proposed scheme incurs a lower cost. We can also see from Fig. \ref{fig:opeCost_s} that as $p$, i.e., the number of pairs sharing a policy, increases, the operating cost decreases. This is because less policies need to be added, introducing less cost.

Fig. \ref{fig:opeCost_l} illustrates the operating cost when $m$ further increases to $1000$. 
We can see that the operating cost of the proposed scheme increases quadratically as $m$ increases. 
Thus, when $m$ exceeds some value, the proposed ABAC scheme will introduce more operating cost than the ACL-based scheme in \cite{zhang2018smart}. 
For example, this value is about $214$ for the case of $n=m$, about $489$ for the case of $n=m/2$ and about $761$ for the case of $n=m/3$.

\begin{figure}[t]
 \centering
 \includegraphics[width=\columnwidth]{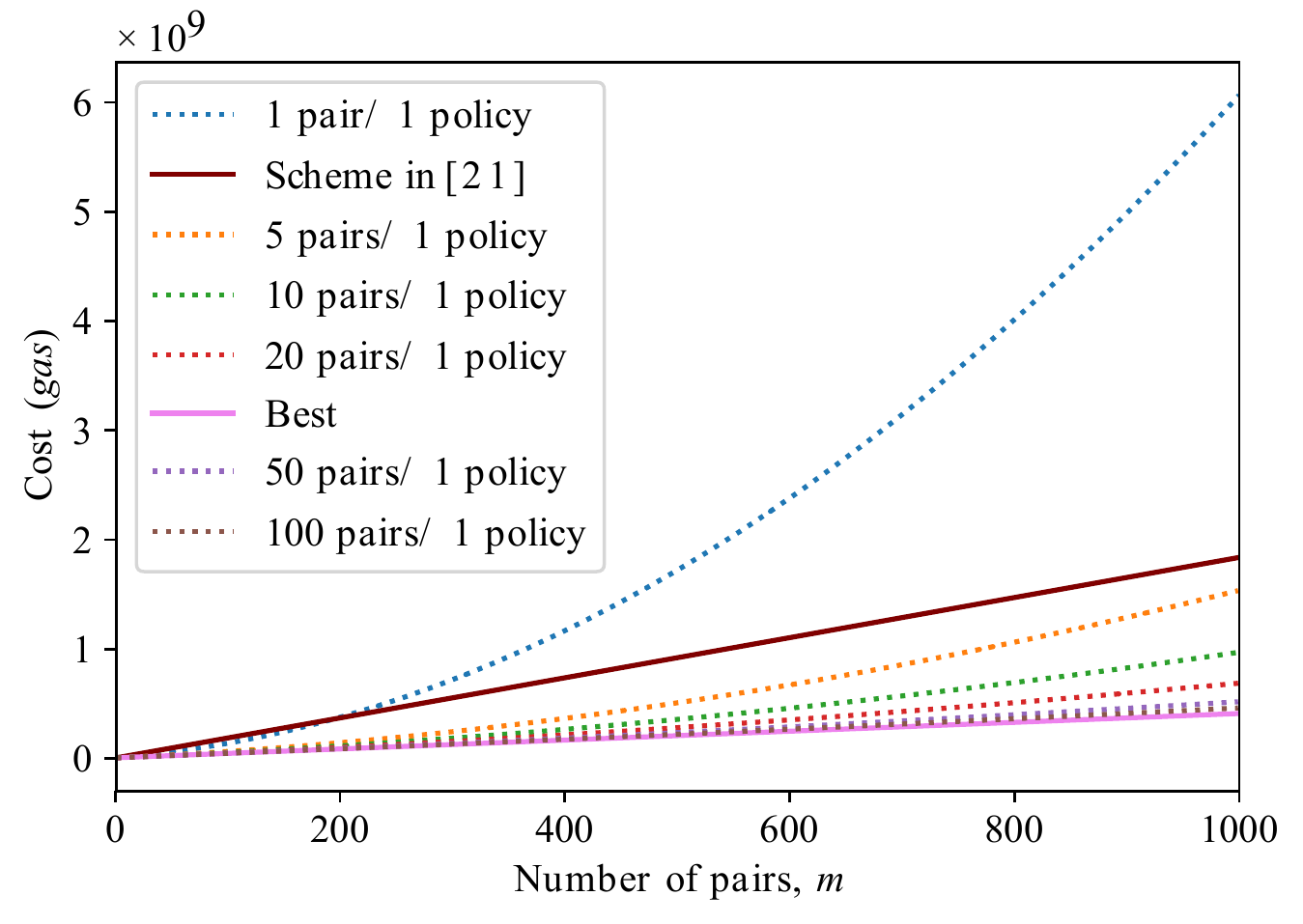}
 \caption{Operating cost when one policy is shared by more pairs.}
 \label{fig:opeCost_b}
\end{figure}

Fig. \ref{fig:opeCost_b} shows the behavior of administrator's operating cost versus the number of subject-object pairs $m$, when each policy is shared by more pairs, i.e., when $p$ is larger. 
As $p$ increases, the cost of our scheme decreases and finally approaches the cost of the best case, i.e., $p=m$.

\subsubsection{Operating Cost in Various Scenarios}
We consider a smart campus containing $1000$ subjects, $150$ objects and $100$ policies. 
We assume each subject randomly and uniformly chooses $15$ objects to access. 
The maximum numbers of subject and object attributes are set to $6$, i.e., $A_s=A_o=6$. 
Also, the maximum numbers of characters in a subject attribute and an object attribute are set to $10$, i.e., $C_s=C_o=10$.
 We consider the following two scenarios of access control in this system.
\begin{itemize}
\item Scenario 1: Adding new objects
\label{sce1}

We consider the scenario of access control on two newly equipped lights in a laboratory of $10$ members including staff and students.
In this case, the attributes of the two lights need to be added. 
If existing policies apply to the new lights, no policies need to be added, which is the best case. 
Otherwise, we need to add new policies as well. 
For simplicity, we only consider the worst case where a new policy is required for each subject-object pair. 
That means the number of policies to add is the product of the number of subjects and the number of the newly-added lights (i.e., $2$).
On the other hand, if the ACL-based scheme in \cite{zhang2018smart} is used, we need to deploy one ACC and policy for each subject-object pair. 
This means, the numbers of policies and ACCs to add are the product of the number of subjects and the number of the newly-added lights.
Table \ref{tb:sce1} shows the operating costs of both schemes, which indicates that the proposed ABAC scheme outperforms the one in \cite{zhang2018smart} in terms of the gas cost.

\begin{table}[t]
 \centering
 \caption{Comparison of operating cost (Scenario 1)}
 \label{tb:sce1}
 \begin{tabular}{|l|r|r|} \hline
  & Gas & USD \\ \hline
  ACL-based Scheme in \cite{zhang2018smart} & 36,701,340 & 38.7567 \\
  Proposed scheme (Best) & 310,136 & 0.32751 \\
  Proposed scheme (Worst) & 16,163,226 & 17.07 \\ \hline
 \end{tabular}
\end{table}

\item Scenario 2: Adding new subjects

This scenario shows the cost when $300$ new members join the university, which corresponds to the case where $300$ subjects are newly added to an IoT system.
First, the attributes of the subjects need to be added. 
Second, If existing policies apply to the new subjects, no policies need to be added, which is the best case. 
Otherwise, we need to add new policies as well. 
For simplicity, we only consider the worst case where a new policy is required for each subject-object pair. 
That means the number of policies to add is the product of the number of newly-added subjects and the number of the objects.
The operating cost of using the scheme in \cite{zhang2018smart}  is also evaluated in this scenario.  
The numbers of policies and ACCs to add are the product of the number of the newly-added subjects and the number of the objects.
Table \ref{tb:sce2} shows the operating costs of both schemes, which indicates that the proposed ABAC scheme consumes less gas than the one in \cite{zhang2018smart} in the best case, while it consumes more gas than the latter in the worst case, which rarely happens in practice.
\end{itemize}

\begin{table}[t]
 \centering
 \caption{Comparison of operating cost (Scenario 2)}
 \label{tb:sce2}
 \begin{tabular}{|l|r|r|} \hline
  & Gas & USD\\ \hline
  ACL-based Scheme in \cite{zhang2018smart} & 8,257,801,500 & 8720.2 \\
  Proposed scheme (Best) & 46,520,400 & 49.125\\
  Proposed scheme (Worst) & 113,438,512,500 & 119791.07 \\ \hline
 \end{tabular}
\end{table}

\subsection{\yet{Discussions}}
\yet{Apart from the monetary cost, another major concern is the throughput issue, i.e., the number of access requests that can be processed per unit time (e.g. second).
The throughput of the proposed ABAC framework depends heavily on the throughput (i.e., number of transactions included in the blockchain per second) of the underlying blockchain systems.
We applied Ethereum 1.0 as the underlying blockchain system in our implementation, the throughput of which is about $15$ transactions per second \cite{ETHthroughput}.
In addition, the access request processing unit ACC in our framework needs to communicate with other contracts through messages, which further introduces latency to the access control process and thus reduces the throughput of the framework.
One of the main reasons behind the low throughput is the consensus algorithm.
Our implementation is based on the popular proof-of-work (PoW) algorithm, which requires a huge amount of calculations to add one block of transactions into the blockchain.
Note that the proposed framework is actually independent of the underlying blockchain systems and consensus algorithms, as long as the blockchain systems support smart contract-like functionality.
This means that the throughput of the proposed framework can be improved by being implemented upon faster blockchains.
One promising solution is Ethereum 2.0, which changes the consensus algorithm from PoW to proof-of-stake (PoS) and adopts the technique of sharding to greatly improve the throughput  performance \cite{ETH2FAQ}.
It is expected that Ethereum 2.0 will enable $64$ to several hundred times more throughput than Ethereum 1.0 \cite{ETH2FAQ}. 
Therefore, one future work is to implement our framework on Ethereum 2.0 to improve its throughput performance.}

%\subsubsection{User Cost}
%
%\begin{figure}[h]
% \centering
% \includegraphics[width=\columnwidth]{fig/usersCost.pdf}
% \caption{User Cost vs. Number of Subject-Object Pairs, $m$}
% \label{fig:user}
%\end{figure}
%
%\begin{figure}[h]
% \centering
% \includegraphics[width=\columnwidth]{fig/usersCost2.pdf}
% \caption{User Cost vs. Number of Policies}
% \label{fig:user2}
%\end{figure}
%
%Users need to spend some money when executing ABIs to perform access control, which is called user cost. 
%We consider a scenario where $m$ pairs of subjects and objects rely on one policy to perform access control. 
%Fig. \ref{fig:user} shows the user cost versus the number of subject-object pairs $m$. We can see that the user cost of both schemes remain constant as $m$ increases, which indicates that the user cost does not depend on the number of subject-object pairs. 
%This is because the user cost depends on the number of policies, which is constant as $m$ increases  in the considered scenarios. 
%We can see also that the proposed ABAC scheme requires more user cost than the ACL-based scheme in \cite{zhang2018smart}.
%
%To show the dependency of user cost of the proposed scheme on the number of policies, Fig. \ref{fig:user2} illustrates the user cost as the number of policies increases. 
%The results indicate that the user cost of the proposed scheme increases as the number of policies increases, because of the increasing cost of executing the \textit{findPolicy()} ABI. 
%On the contrary, the user cost of the scheme in  \cite{zhang2018smart} remains unchanged as the number of policies increases.

\section{Conclusion}\label{sec:conclusion}
In this paper, we proposed an ABAC framework for smart cities by using Ethereum smart contracts to manage ABAC policies, attributes of subjects and objects and perform access control. 
A local private Ethereum blockchain network was constructed to implement the proposed framework and demonstrate its feasibility. 
Based on the implementation, we also conducted extensive experiments to evaluate the monetary cost of the proposed scheme in terms of gas consumption. 
First, we evaluated the gas consumption of basic operations provided by the framework, like access control and the management of  subjects, objects and policies. 
Second, we evaluated the gas consumed by the system administrators for deploying the proposed framework on the blockchain (i.e., deployment cost) and running the framework after deployment (i.e., operating cost). 
For comparison, we also evaluated the initial cost and operating cost of an existing  ACL-based framework. 
The experiment results showed that although our framework introduces a larger deployment cost than the ACL-based framework, it introduces less operating cost in general, especially for IoT systems containing a large number of subjects and objects with common attributes. Smart cities are typical examples of such systems. 
\yet{Although the prototype demonstrates the feasibility of the proposed framework, it can hardly reflect the performance of the framework in large-scale smart cities. 
Thus, we will consider the implementation of the framework in environments with larger scales as our future work.}
%We also evaluated the cost of users of these two schemes.
%The results showed that our scheme incurs larger user cost than the ACL-based scheme. 

\section*{Acknowledgment}
This work was supported in part by the Japan Society for the Promotion of Science (JSPS)
KAKENHI (A) under Grant 19H01103, the Telecommunications Advancement Foundation, and the Support
Center for Advanced Telecommunications (SCAT) Technology Research Foundation.

\bibliographystyle{IEEEtran}
\bibliography{Manuscript.bib}

\begin{thebibliography}{10}
\providecommand{\url}[1]{#1}
\csname url@samestyle\endcsname
\providecommand{\newblock}{\relax}
\providecommand{\bibinfo}[2]{#2}
\providecommand{\BIBentrySTDinterwordspacing}{\spaceskip=0pt\relax}
\providecommand{\BIBentryALTinterwordstretchfactor}{4}
\providecommand{\BIBentryALTinterwordspacing}{\spaceskip=\fontdimen2\font plus
\BIBentryALTinterwordstretchfactor\fontdimen3\font minus
  \fontdimen4\font\relax}
\providecommand{\BIBforeignlanguage}[2]{{%
\expandafter\ifx\csname l@#1\endcsname\relax
\typeout{** WARNING: IEEEtran.bst: No hyphenation pattern has been}%
\typeout{** loaded for the language `#1'. Using the pattern for}%
\typeout{** the default language instead.}%
\else
\language=\csname l@#1\endcsname
\fi
#2}}
\providecommand{\BIBdecl}{\relax}
\BIBdecl

\bibitem{yutaka19using}
M.~Yutaka, Y.~Zhang, M.~Sasabe, and S.~Kasahara, ``{Using ethereum blockchain
  for distributed attribute-based access control in the internet of things},''
  in \emph{Proc. of IEEE GLOBECOM 2019}, 2019.

\bibitem{gubbi2013iot}
J.~Gubbi, R.~Buyya, S.~Marusic, and M.~Palaniswami, ``{Internet of things
  (IoT): A vision, architectural elements, and future directions},''
  \emph{Future Generation Computer Systems}, vol.~29, no.~7, pp. 1645--1660,
  2013.

\bibitem{alaba2017iot}
F.~A. Alaba, M.~Othman, I.~A.~T. Hashem, and F.~Alotaibi, ``Internet of things
  security: A survey,'' \emph{Journal of Network and Computer Applications},
  vol.~88, pp. 10 -- 28, 2017.

\bibitem{Mirai}
``{Mirai botnet linked to dyn DNS DDoS attacks},'' available at
  \url{https://www.flashpoint-intel.com/blog/cybercrime/mirai-botnet-linked-dyn-dns-ddos-attacks/}.

\bibitem{webcam}
``{Breached webcam and baby monitor site flagged by watchdogs},'' available at
  \url{https://www.bbc.com/news/technology-30121159}.

\bibitem{Sookhak2019IEEECST}
M.~{Sookhak}, H.~{Tang}, Y.~{He}, and F.~R. {Yu}, ``Security and privacy of
  smart cities: A survey, research issues and challenges,'' \emph{IEEE
  Communications Surveys and Tutorials}, vol.~21, no.~2, pp. 1718--1743, 2019.

\bibitem{Eckhoff2018IEEECST}
D.~{Eckhoff} and I.~{Wagner}, ``Privacy in the smart city—applications,
  technologies, challenges, and solutions,'' \emph{IEEE Communications Surveys
  and Tutorials}, vol.~20, no.~1, pp. 489--516, 2018.

\bibitem{Algarni2019IEEEAccess}
A.~{Algarni}, ``A survey and classification of security and privacy research in
  smart healthcare systems,'' \emph{IEEE Access}, vol.~7, pp.
  101\,879--101\,894, 2019.

\bibitem{locken}
``{Smart access control is essential to the future of Smart Cities},''
  available at
  \url{https://www.locken.eu/smart-access-control-is-essential-to-the-future-of-smart-cities/}.

\bibitem{Gharaibeh2017IEEECST}
A.~{Gharaibeh}, M.~A. {Salahuddin}, S.~J. {Hussini}, A.~{Khreishah},
  I.~{Khalil}, M.~{Guizani}, and A.~{Al-Fuqaha}, ``Smart cities: A survey on
  data management, security, and enabling technologies,'' \emph{IEEE
  Communications Surveys and Tutorials}, vol.~19, no.~4, pp. 2456--2501, 2017.

\bibitem{yavari2017scalable}
A.~Yavari, A.~S. Panah, D.~Georgakopoulos, P.~P. Jayaraman, and R.~V. Schyndel,
  ``{Scalable role-based data disclosure control for the internet of things},''
  in \emph{Proc. of 2017 IEEE 37th International Conference on Distributed
  Computing Systems}, 2017, pp. 2226--2233.

\bibitem{liu2017anaccess}
Q.~Liu, H.~Zhang, J.~Wan, and X.~Chen, ``{An access control model for resource
  sharing based on the role-based access control intended for multi-domain
  manufacturing internet of things},'' \emph{IEEE Access}, vol.~5, no.~2, pp.
  7001--7011, 2017.

\bibitem{yuan2005attributed}
E.~Yuan and J.~Tong, ``{Attributed based access control (ABAC) for web
  services},'' in \emph{Proc. of IEEE International Conference on Web
  Services}, 2005, pp. 561--569.

\bibitem{hwang2005TKDE}
{Hua Wang}, {Jinli Cao}, and {Yanchun Zhang}, ``A flexible payment scheme and
  its role-based access control,'' \emph{IEEE Transactions on Knowledge and
  Data Engineering}, vol.~17, no.~3, pp. 425--436, 2005.

\bibitem{hwang2009TKDE}
H.~{Wang}, Y.~{Zhang}, and J.~{Cao}, ``Effective collaboration with information
  sharing in virtual universities,'' \emph{IEEE Transactions on Knowledge and
  Data Engineering}, vol.~21, no.~6, pp. 840--853, 2009.

\bibitem{li2011privacy}
M.~Li, X.~Sun, H.~Wang, Y.~Zhang, and J.~Zhang, ``Privacy-aware access control
  with trust management in web service,'' \emph{World Wide Web}, vol.~14,
  no.~4, pp. 407--430, 2011.

\bibitem{Hernandez-Ramos2013}
J.~L. Hernandez-Ramos, A.~J. Jara, L.~Marin, and A.~F. Skarmeta, ``{Distributed
  capability-based access control for the internet of things},'' \emph{Journal
  of Internet Services and Information Security}, vol.~3, no. 3/4, pp. 1--16,
  2013.

\bibitem{Sciancalepore2018}
S.~Sciancalepore, G.~Piro, D.~Caldarola, G.~Boggia, and G.~Bianchi, ``{On the
  design of a decentralized and multi-authority access control scheme in
  federated and cloud-assisted cyber-physical systems},'' \emph{IEEE Internet
  of Things Journal}, vol.~5, no.~6, pp. 5190--5204, 2018.

\bibitem{dukkipati2018decentralized}
C.~Dukkipati, Y.~Zhang, and L.~C. Cheng, ``{Decentralized, blockchain based
  access control framework for the heterogeneous internet of things},'' in
  \emph{Proc. of 3rd Workshop on Attribute Based Access Control}, 2018, pp.
  61--69.

\bibitem{wang2019anattribute}
P.~Wang, Y.~Yue, W.~Sun, and J.~Liu, ``{An attribute-based distributed access
  control for blockchain-enabled IoT},'' in \emph{2019 International Conference
  on Wireless and Mobile Computing, Networking and Communications (WiMob)},
  2019, pp. 1--6.

\bibitem{maesa2019ablockchain}
D.~D.~F. Maesa, P.~Mori, and L.~Ricci, ``A blockchain based approach for the
  definition of auditable access control systems,'' \emph{Computers \&
  Security}, vol.~84, pp. 93 -- 119, 2019.

\bibitem{dorri2017blockchain}
A.~Dorri, S.~S. Kanhere, R.~Jurdak, and P.~Gauravaram, ``{Blockchain for IoT
  security and privacy: The case study of a smart home},'' in \emph{Proc. of
  IEEE PerCom Workshops}, 2017, pp. 618--623.

\bibitem{maesa2017blockchain}
D.~F. Maesa, P.~Mori, and L.~Ricci, ``{Blockchain based access control},'' in
  \emph{Proc. of IFIP International Conference on Distributed Applications and
  Interoperable Systems}, 2017, pp. 206--220.

\bibitem{zhu2018tbac}
Y.~Zhu, Y.~Qin, G.~Gan, S.~Yang, and W.~C.-C. Chu, ``{TBAC: Transaction-based
  access control on blockchain for resource sharing with cryptographically
  decentralized authorization},'' in \emph{Proc. of 2018 42nd IEEE
  International Conference on Computer Software \& Applications}, 2018, pp.
  535--544.

\bibitem{ouaddah2017access}
A.~Ouaddah, H.~Mousannif, A.~A. Elkalam, and A.~A. Ouahman, ``{Access control
  in the internet of things: Big challenges and new opportunities},''
  \emph{Computer Networks}, vol. 112, pp. 237--262, 2017.

\bibitem{zhang2018smart}
Y.~Zhang, S.~Kasahara, Y.~Shen, X.~Jiang, and J.~Wan, ``{Smart contract-based
  access control for the internet of things},'' \emph{IEEE Internet of Things
  Journal}, vol.~6, no.~2, pp. 1594--1605, 2019.

\bibitem{tanzeela2020data}
T.~Sultana, A.~Ghaffar, M.~Azeem, Z.~Abubaker, M.~U. Gurmani, and N.~Javaid,
  ``{Data sharing system integrating access control based on smart contracts
  for IoT},'' in \emph{International Conference on P2P, Parallel, Grid, Cloud
  and Internet Computing}.\hskip 1em plus 0.5em minus 0.4em\relax Springer,
  2019, pp. 863--874.

\bibitem{xu2018blendcac}
R.~Xu, Y.~Chen, E.~Blasch, and G.~Chen, ``{BlendCAC: A smart contract enabled
  decentralized capability-based access control mechanism for the IoT},''
  \emph{Computers}, vol.~7, no.~3, pp. 39--65, 2018.

\bibitem{nakamura19capbac}
Y.~Nakamura, Y.~Zhang, M.~Sasabe, and S.~Kasahara, ``{Capability-based access
  control for the internet of things: An ethereum blockchain-based scheme},''
  in \emph{Proc. of IEEE GLOBECOM 2019}, 2019.

\bibitem{albreiki2019decentralized}
H.~Albreiki, L.~Alqassem, K.~Salah, M.~H. Rehman, and D.~Svetinovic,
  ``{Decentralized access control for IoT data using blockchain and trusted
  oracles},'' in \emph{In Proceedings of IEEE International Conference on
  Industrial Internet (ICII)}, Nov. 2019, pp. 248--257.

\bibitem{lyu2020sbac}
Q.~Lyu, Y.~Qi, X.~Zhang, H.~Liu, Q.~Wang, and N.~Zheng, ``Sbac: A secure
  blockchain-based access control framework for information-centric
  networking,'' \emph{Journal of Network and Computer Applications}, vol. 149,
  p. 102444, 2020.

\bibitem{jasonpaul2018rbacsc}
J.~P. Cruz, Y.~Kaji, and N.~Yanai, ``{RBAC-SC: Role-based access control using
  smart contract},'' \emph{IEEE Access}, vol.~6, pp. 12\,240--12\,251, Mar.
  2018.

\bibitem{hao2019multi}
G.~Hao, E.~Meamari, and C.-C. Shen, ``{Multi-authority attribute-based access
  control with smart contract},'' in \emph{Proc. of 2019 International
  Conference on Blockchain Technology}, 2019, pp. 6--11.

\bibitem{ding2019anovel}
S.~Ding, J.~Cao, C.~Li, K.~Fan, and H.~Li, ``{A novel attribute-based access
  control scheme using blockchain for IoT},'' \emph{IEEE Access}, vol.~7, pp.
  38\,431--38\,441, 2019.

\bibitem{yu2020enabling}
G.~Yu, X.~Zha, X.~Wang, W.~Ni, K.~Yu, P.~Yu, J.~A. Zhang, R.~P. Liu, and Y.~J.
  Guo, ``{Enabling attribute revocation for fine-grained access control in
  blockchain-IoT systems},'' \emph{IEEE Transactions on Engineering
  Management}, 2020.

\bibitem{suciu2019attribute}
G.~Suciu, C.-I. Istrate, A.~Vulpe, M.-A. Sachian, M.~Vochin, A.~Farao, and
  C.~Xenakis, ``Attribute-based access control for secure and resilient smart
  grids,'' in \emph{6th International Symposium for ICS \& SCADA Cyber Security
  Research 2019 6}, 2019, pp. 67--73.

\bibitem{Bitcoin}
``{Bitcoin - open source p2p money},'' available at
  \url{https://bitcoin.org/en/ }.

\bibitem{Ethereum}
``{An introduction to Ethereum platform},'' available at
  \url{http://ethdocs.org/en/latest/introduction/what-is-ethereum.html}.

\bibitem{standard2013extensible}
``Extensible access control markup language (xacml) version 3.0,'' available at
  \url{http://docs. oasis-open.
  org/xacml/2.0/access_control-xacml-2.0-core-spec-os. pdf}.

\bibitem{SC}
``{A next-generation smart contract and decentralized application platform},''
  available at
  \url{https://cryptorating.eu/whitepapers/Ethereum/Ethereum_white_paper.pdf}.

\bibitem{geth}
``{Geth client for building private blockchain networks.}'' available at
  \url{https://github.com/ethereum/go-ethereum/wiki/geth}.

\bibitem{Remix}
``{Remix IDE for ethereum smart contract programming.}'' available at
  \url{https://remix.ethereum.org/}.

\bibitem{web3}
``{Web3 javascript api to interact with Ethereum nodes.}'' available at
  \url{https://github.com/ethereum/wiki/wiki/JavaScript-API}.

\bibitem{EthGas}
``{ETH gas station},'' available at
  \url{https://ethgasstation.info/calculatorTxV.php}.

\bibitem{ETHthroughput}
``{On sharding blockchains FAQs},'' available at
  \url{https://eth.wiki/sharding/Sharding-FAQs}.

\bibitem{ETH2FAQ}
``{What is the difference between Ethereum 1.0 and Ethereum 2.0?}'' available
  at \url{https://consensys.net/knowledge-base/ethereum-2/faq/}.

\end{thebibliography}

\end{document}